\pdfoutput=1
\documentclass[sigplan,nonacm]{acmart}

\usepackage{tabularx}
\usepackage{xltabular} 

\newcommand{\tool}{\textsc{VSpector}}

\newcommand{\inout}{{\itshape in\,\textbar\,out}}
\newcolumntype{P}{r@{\,\textbar\,}l}
\newcommand{\hd}[2]{\begin{tabular}[c]{@{}c@{}}#1\\ #2\end{tabular}}
\newcommand{\nbugs}{42}         
\newcommand{\nbugsall}{73}      
\newcommand{\ndup}{31}          
\newcommand{\nfixed}{19}        
\newcommand{\nconfirmed}{11}     
         
\newcommand{\nrules}{5{,}685}   
\newcommand{\ncand}{1{,}083}    
\newcommand{\naudited}{217}     
\newcommand{\ntrue}{148}        
\newcommand{\nprecall}{68.2}    
\newcommand{\nprecnone}{13.7}   
\newcommand{\nprecspec}{38.5}   
\newcommand{\nprecode}{29.6}    
\newcommand{\nspeckept}{384}    
\newcommand{\ncodekept}{500}    
\newcommand{\nrejected}{866}    
\newcommand{\nfuzzh}{24}        
\newcommand{\locfilelines}{1{,}708} 
\newcommand{\locsliceblocks}{ten}   
\newcommand{\locslicelines}{227}

\title{\tool: Specification-Driven Bug Detection for RISC-V CPUs}

\author{Tianyu Jia}
\affiliation{
  \institution{Renmin University of China}
  \city{Beijing}
  \country{China}
}
\email{2025103980@ruc.edu.cn}

\author{Zhaoyang Yu}
\affiliation{
  \institution{Renmin University of China}
  \city{Beijing}
  \country{China}
}
\email{2023202273@ruc.edu.cn}

\author{Yuanliang Chen}
\affiliation{
  \institution{Renmin University of China}
  \city{Beijing}
  \country{China}
}
\email{sard.chen@gmail.com}

\author{Wei You}
\affiliation{
  \institution{Renmin University of China}
  \city{Beijing}
  \country{China}
}
\email{youwei@ruc.edu.cn}

\author{Jianjun Huang}
\affiliation{
  \institution{Renmin University of China}
  \city{Beijing}
  \country{China}
}
\email{hjj@ruc.edu.cn}

\author{Bin Liang}
\authornote{Corresponding author.}
\affiliation{
  \institution{Renmin University of China}
  \city{Beijing}
  \country{China}
}
\email{liangb@ruc.edu.cn}

\renewcommand{\shortauthors}{Jia et al.}

\begin{document}

\begin{abstract}
Detecting RTL design bugs in open-source RISC-V CPU implementations is critical for ensuring system reliability. Traditional detection approaches inherently rely on predefined artifacts. In this paper, we leverage the official, natural-language RISC-V specifications as an effective information source for bug detection. We present \tool{}, a specification-driven bug detection pipeline that directly checks whether CPU register-transfer level (RTL) implementations adhere to official specification rules, without requiring specialized construction of reference models, formal properties, or custom bug patterns. To resolve the key technical trade-off between broad context scope and model reasoning accuracy when using Large Language Models (LLMs), \tool{} employs a stepwise context refinement scheme across a four-stage pipeline: rule extraction, implementation localization, candidate identification, and sequential violation auditing. We evaluate \tool{} on two industrial-strength RISC-V CPUs, CVA6 and XiangShan. Out of \naudited{} reported candidates, manual inspection confirmed \ntrue{} true violations, representing a \nprecall{}\% precision. These violations correspond to \nbugsall{} distinct bugs, including \textbf{\nbugs{}} previously unknown bugs. In our comparative experiments, DiveFuzz, a state-of-the-art CPU fuzzer, detected none of these new bugs during \nfuzzh{}-hour runs per CPU. All \nbugs{} new bugs have been reported upstream, with developers already fixing \nfixed{} and confirming an additional \nconfirmed{} (\textbf{30} in total), demonstrating that specification-driven auditing is a practical and complementary strategy for CPU bug detection.
\end{abstract}

\maketitle
\hypersetup{pdfauthor={Tianyu Jia, Zhaoyang Yu, Yuanliang Chen, Wei You, Jianjun Huang, Bin Liang}}

\section{Introduction}
\label{sec:intro}

RISC-V is an open-standard instruction set architecture (ISA) used in systems ranging from embedded devices to large-scale computing systems. In particular, open-source RISC-V CPUs provide accessible register-transfer level (RTL) designs. For instance, representative designs such as CVA6~\cite{cva6} and XiangShan~\cite{xiangshan} make their full designs available in SystemVerilog~\cite{systemverilog} and Chisel~\cite{chisel}, holding great promise for widespread adoption.

The CPU is the core of a computer system. Like software, CPUs are susceptible to bugs that affect overall system operation. Unlike software, however, CPUs are hard to be modified once manufactured. Post-silicon bugs can usually only be mitigated by disabling affected features~\cite{riscover}, which often leads to significant performance drops. Therefore, detecting bugs in CPU designs is critical.

A variety of approaches have been proposed for detecting bugs in RISC-V CPUs, successfully uncovering some real bugs. CPU fuzzing~\cite{difuzzrtl,cascade,divefuzz} feeds identical input vectors into both the target CPU design and a reference ISA simulator, identifying potential bugs by checking for discrepancies in architectural states such as registers and memory. Formal verification~\cite{riscvformal} aims to prove whether a CPU's RTL design satisfies predefined correctness properties. Static analysis~\cite{qihe}, borrowing concepts from software analysis, checks RTL designs against summarized bug patterns, making it scalable to large-scale designs.

However, the effectiveness of these existing methods inherently relies on the artifacts they employ, namely reference models, properties, and bug patterns. In practice, these artifacts are not always reliable or comprehensive. Reference models themselves may suffer from implementation issues (e.g., as shown in Section~\ref{sec:motivate-existing}, simplifications in the widely used ISA simulators NEMU~\cite{nemu} and Spike~\cite{spike} prevent them from detecting certain bugs), whereas manually crafted properties and patterns are prone to human error and incompleteness. This raises a key research question: \textit{Can we find alternative, high-quality information sources beyond conventional artifacts to identify CPU bugs?}

In this study, we identify the official RISC-V specifications~\cite{riscvunpriv,riscvpriv,riscvaia,riscvdebug}, maintained by RISC-V International, as a valuable yet underutilized information source for constructing detection artifacts. Independent of specific CPU implementations, these specifications comprehensively define the rules in natural language that compliant CPU designs must obey, including instruction semantics, privilege modes, control registers, interrupts, address translation, and external debugging. These documents are structured into distinct sections, each of which typically states several related rules. In practice, a CPU's RTL design that fails to adhere to a required rule is highly likely to manifest as a bug, which we term a \textit{specification violation}.

Section~\ref{sec:motivate} illustrates such a violation in XiangShan, which violates a rule regarding the hypervisor memory-management fence: the fence must ignore the reserved bits of its operands. However, the corresponding Chisel implementation lets two of these reserved bits take part in the comparison that selects translations to flush.

\begin{figure}[H]
\centering
\includegraphics[width=\columnwidth]{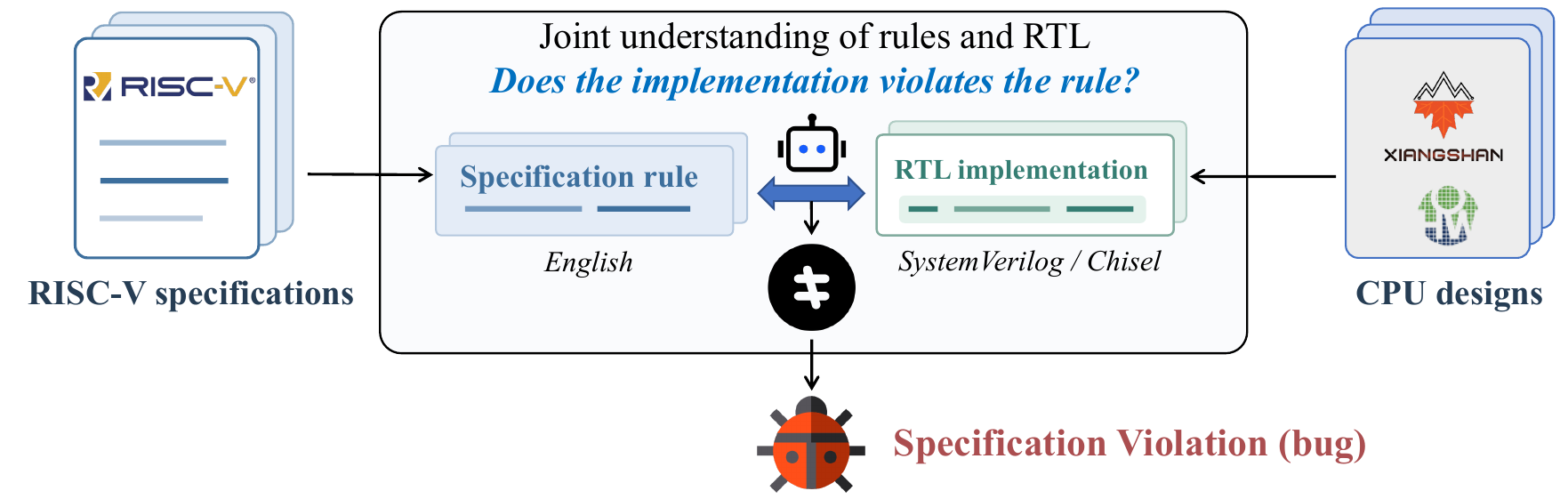}
\caption{Specification violation detection.}
\label{fig:intro}
\end{figure}

Identifying specification violations requires a joint understanding of natural-language specifications and RTL CPU designs. Recent advances in Large Language Models (LLMs) provide a viable path to perform such cross-domain analysis. As illustrated in \autoref{fig:intro}, we can extract explicit rules from the specification and derive their corresponding RTL implementations from the CPU design, and then leverage LLMs to check whether the implementation adheres to the specified rules, eliminating the need for specialized construction of reference models, formal properties, or custom bug patterns.

However, realizing such specification-to-code alignment introduces a key technical challenge. Concretely, performing the cross-domain analysis faces a clear trade-off between context scope and reasoning accuracy. On the one hand, a single specification section often contains multiple rules alongside scattered descriptions, with potential dependencies between rules. Similarly, the corresponding RTL implementation can be spread across multiple modules. Feeding intertwined rules and broad RTL implementations directly into LLMs reduces the model's attention on key logic, lowering detection accuracy. On the other hand, overly narrowing the context scope causes the LLM to miss necessary specification or implementation details, leading to false positives.

To address this challenge, we adopt a stepwise refinement scheme and propose a multi-stage pipeline \textbf{\tool{}}. It combines localized focusing with dynamic context detection and auditing across four stages: (1) \textbf{Rule extraction} extracts targeted specification rules section by section, preserving applicability conditions and source locations so that each rule can be checked independently; (2) \textbf{Implementation localization} searches the CPU repository for code implementing each rule, restricting inspection to the relevant lines; (3) \textbf{Candidate identification} checks whether the localized code violates the rule, triggering a ReAct process to retrieve supplementary RTL code on demand whenever context is insufficient; and (4) \textbf{Violation audit} dynamically retrieves supplementary specification clauses and RTL context for each candidate, performing sequential specification-side and code-side reviews to effectively mitigate false positives.

We evaluate \tool{} on CVA6 and XiangShan. Among the \naudited{} candidate violations reported by \tool{}, manual inspection confirmed \ntrue{} as true violations, achieving a precision of \nprecall\%. These violations correspond to \nbugsall{} distinct bugs, \nbugs{} of which were previously unknown. All \nbugs{} new bugs have been reported upstream, with developers already fixing \nfixed{} and confirming an additional \nconfirmed{}. Notably, DiveFuzz~\cite{divefuzz}, a state-of-the-art (SOTA) RISC-V CPU fuzzer, failed to detect any of these \nbugs{} bugs during a \nfuzzh{}-hour run on the same commits. 
These results demonstrate that directly checking RTL implementations against RISC-V specification rules is a practical and effective strategy for uncovering real-world CPU bugs.

In summary, our main contributions are as follows:

\begin{itemize}

\item \textbf{New Scheme:} We introduce a new specification-driven bug detection scheme for open-source RISC-V CPUs. By treating official specification documents directly as detection artifacts, our scheme uncovers bugs by identifying inconsistencies between hardware specification intent and RTL implementations, without requiring predefined reference models, formal properties, or bug patterns.
    
\item \textbf{Multi-stage Pipeline:} We design \tool{}, a multi-stage pipeline based on a stepwise refinement scheme that integrates localized focusing with dynamic context retrieval. It initially restricts analysis to an individual rule and its core RTL code, and then progressively retrieves additional context on demand to conduct bug detection and rigorous auditing, ultimately mitigating false positives.
    
\item \textbf{Detection Results:} We demonstrate the effectiveness of \tool{} on two industrial-strength RISC-V CPUs, CVA6 and XiangShan. \tool{} achieved a precision of \nprecall\% and uncovered \nbugsall{} distinct bugs (\textbf{\nbugs{}} previously unknown), demonstrating complementary bug-detection capability to SOTA CPU fuzzing.
All new bugs were reported upstream, with developers already fixing \nfixed{} and confirming an additional \nconfirmed{} (\textbf{30} in total).

\end{itemize}

\section{Motivating Examples}
\label{sec:motivate}

\begin{figure}[t]
\centering
\includegraphics[width=\columnwidth]{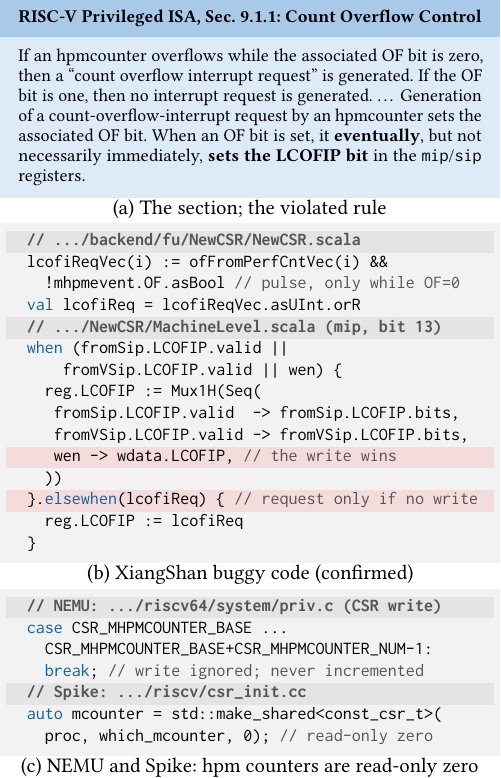}
\caption{A \emph{specification violation} that \tool{} found in XiangShan, confirmed upstream: (a)~the rule; (b)~the RTL, where a same-cycle write to \texttt{mip} overrides the overflow request; (c)~NEMU and Spike, whose counters never overflow.}
\label{fig:lcofi}
\end{figure}

\begin{figure}[t]
\centering
\includegraphics[width=\columnwidth]{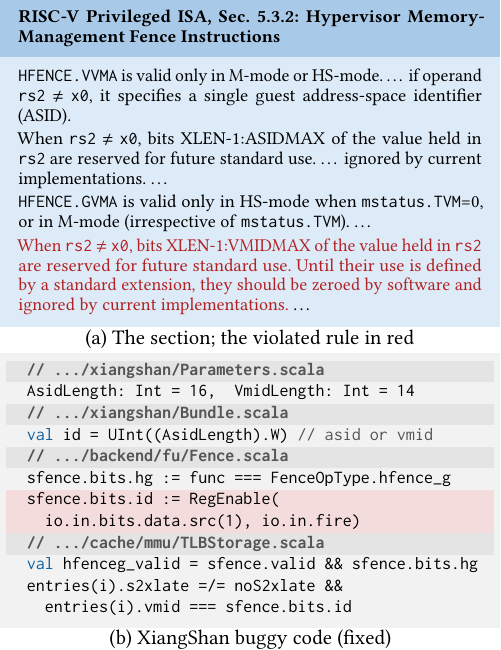}
\caption{A \emph{specification violation} that \tool{} found in XiangShan, since fixed upstream: (a)~the section, the violated rule in red; (b)~the RTL that carries two reserved bits of \texttt{rs2} into the VMID comparison.}
\label{fig:hfence}
\end{figure}

We use two violations that \tool{} found in XiangShan: a count-overflow interrupt that the CPU drops (\autoref{fig:lcofi}), and the \texttt{HFENCE.GVMA} of \autoref{fig:hfence}, which, when either of two reserved bits of its \texttt{rs2} operand is set, matches no TLB entry and leaves the stale translations in place. Section~\ref{sec:motivate-existing} discusses the challenges these violations pose for CPU fuzzing and static analysis; Section~\ref{sec:motivate-takes} shows what detecting them from the specification takes.

When a hardware performance counter overflows, the OF bit of its \texttt{mhpmevent} register is set and, by the rule in \autoref{fig:lcofi}(a), the LCOFIP bit of \texttt{mip}, the pending bit of the count-overflow interrupt, must eventually be set as well. XiangShan raises the request to \texttt{mip} as a one-cycle pulse while OF is still zero. A software write to \texttt{mip} or \texttt{sip} in the same cycle takes priority and drops the pulse (\autoref{fig:lcofi}(b)), even a write that leaves LCOFIP alone, such as clearing the software-interrupt bit. OF stays set and blocks any further request from the counter, so the interrupt is lost.

An \texttt{HFENCE.GVMA} whose register operand \texttt{rs2} is not \texttt{x0} flushes only the guest translations whose virtual machine identifier (VMID) equals the VMID given in \texttt{rs2}, and the rule in \autoref{fig:hfence}(a) requires the reserved bits of \texttt{rs2} above the VMID to be ignored. XiangShan compares \texttt{rs2}, carried in a 16-bit fence identifier, against the 14-bit VMID stored in each TLB entry, so the two reserved bits just above the VMID take part (\autoref{fig:hfence}(b)). A fence whose \texttt{rs2} sets either bit matches no entry and leaves the stale translations in the TLB.

\begin{figure*}[t]
\centering
\includegraphics[width=\textwidth]{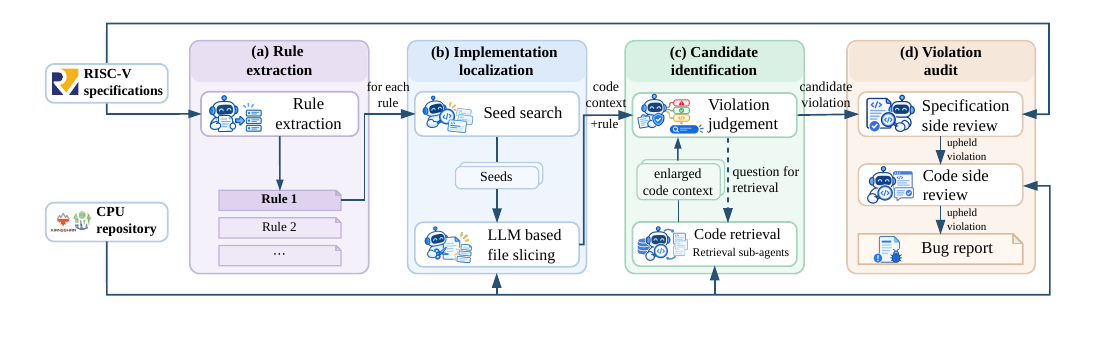}
\caption{Overview of \tool{}. Rule extraction~(a) runs once per specification section; implementation localization~(b) and candidate identification~(c) run once per rule; the violation audit~(d) runs once per candidate violation. Boxes with a robot icon are LLM calls; the stages in (b)--(d) search and read the CPU repository.}
\label{fig:overview}
\end{figure*}

\subsection{Challenges for Existing Approaches}
\label{sec:motivate-existing}

CPU fuzzing~\cite{difuzzrtl,cascade,divefuzz} runs generated programs on the RTL under test and on an ISA simulator that serves as the reference model, and reports a bug when their architectural states diverge. A bug is found only when the reference model implements the behavior correctly and a program triggers it. The count-overflow violation fails the first condition. NEMU~\cite{nemu} and Spike~\cite{spike} simplify the performance counters to read-only zero, which the specification permits (\autoref{fig:lcofi}(c)), so comparing LCOFIP against them would flag every correct overflow; DiffTest~\cite{difftest,xiangshan}, XiangShan's differential testing framework, therefore copies the bit from the RTL into NEMU, and the comparison carries no expectation that an overflow must set it. For the same-cycle write of \autoref{fig:lcofi}(b), both sides read zero. The \texttt{HFENCE.GVMA} violation fails the second. The violation surfaces only after a program enables two-stage address translation, brings a guest translation into the TLB, changes the page-table entry behind it, issues an \texttt{HFENCE.GVMA} with one of the two bits set, and touches the page again; \nfuzzh{} hours of DiveFuzz did not find it (Section~\ref{sec:eval-fuzz}).

Static analysis~\cite{qihe,cweat,chisa} needs neither a program nor a reference model. It searches the RTL for bug patterns fixed in advance, such as a register without reset, an unreachable state of a finite-state machine, or a bit-width mismatch that truncates a value~\cite{qihe,cweat}, or it checks the assertions that the developers wrote into the code~\cite{chisa}. Detecting specification violations with these approaches depends on whether the relevant behavior is covered by the predefined patterns or developer-written assertions. Generic bug patterns often do not encode specification-specific semantics, limiting their ability to detect violations that depend on such knowledge.

\subsection{What Detection from the Specification Takes}
\label{sec:motivate-takes}

A section of the specification states several rules at once. The section in \autoref{fig:hfence}(a) defines both \texttt{HFENCE.VVMA} and \texttt{HFENCE.GVMA}, with the privilege modes, ordering, and operands of each; \tool{} extracts ten rules from it. The violated rule is one sentence about \texttt{rs2} of \texttt{HFENCE.GVMA}; an earlier sentence in the same section says the same about \texttt{rs2} of \texttt{HFENCE.VVMA}, with the ASID in place of the VMID. Checking the whole section at once spreads attention over all of these rules; extracting them and checking each on its own keeps the attention on one.

Deciding whether the RTL violates a rule takes the meaning of the code, and the code that implements a rule is spread over several files. Seeing the violation in \autoref{fig:hfence}(b) takes four files together: the parameters give the ASID and the VMID different widths; the bundle gives the shared fence identifier the ASID width; the fence unit assigns \texttt{rs2} to it, which in Chisel truncates the value, so the upper reserved bits are ignored as the rule requires; and the TLB compares the identifier against the stored VMID, which zero-extends the VMID, so the two bits between the widths take part. The fence unit alone shows \texttt{rs2} assigned to an identifier whose width and use lie elsewhere, so the reserved bits look either fully forwarded or fully ignored; the decision comes out right only with all four files gathered from across the repository and read together.

\section{Design}
\label{sec:design}
    
\subsection{Overview}
\label{sec:overview}

\tool{} takes the RISC-V specification documents and the RTL repository of a CPU as the input and checks the CPU against the specifications. 
\autoref{fig:overview} shows the workflow of \tool{}, involving four stages. 
(a) Given a specification document, \tool{} extracts from it the individual rules that must be implemented by the CPU. 
(b) For each rule, \tool{} leverages the LLM to understand it and searches for the corresponding code snippets as the seeds, which then act as the slicing criteria for extracting the correlated implementation logic to form the code context of the rule. 
(c) With the rule and its code context, the LLM judges whether the rule may be violated. If it needs more code context to make the decision, sub-agents are launched to perform the code retrieval and enlarge the code context. 
(d) The candidate violations are passed to an additional audit against the full specification and the code repository,  to eliminate the false warnings. 
The remaining are reported as the final violations.

\subsection{Rule Extraction}
\label{sec:extract}
    
\begin{figure}[t]
\centering
\includegraphics[width=\columnwidth]{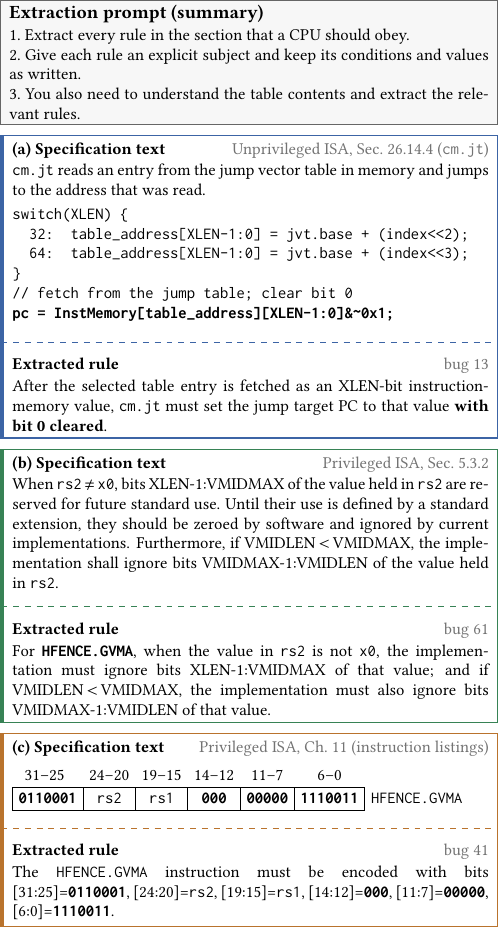}
\caption{The extraction prompt (top) and three pieces of specification
text with the rule extracted from each: (a)~pseudocode, (b)~a sentence
that does not name its instruction, (c)~a row of an encoding table.
The bug index beside each rule corresponds to the row of Table~\ref{tab:bugs} of the
bug it surfaced.}
\label{fig:rules}
\end{figure}
    
A \emph{rule} is a statement in the RISC-V specifications of the correct behavior of a CPU. For example, a rule can indicate how an instruction behaves when executed in a particular CPU state, what value a CSR takes on reset and how it is updated, or which exception is raised in a given situation.
    
To cover as many of these rules as possible, \tool{} does not rely on keywords such as \emph{must}, \emph{shall} and \emph{if}. Instead, it splits each document at the section headings and feeds the LLM with the sections one by one. The LLM extracts the rules from each given section. The simplified prompt is shown at the top of \autoref{fig:rules}, 
which asks the LLM to extract
what a CPU should obey,
which may be described by an individual statement, continuous statements with tight semantic connections, a snippet of sample code and/or a table for the register bit setting. 
Finally, the LLM returns the rules found in the section as a list.

\autoref{fig:rules} shows three example rules
extracted from the text of different forms. In~(a), the behavior of \texttt{cm.jt} is given as a piece of pseudocode, and the rule is depicted in its last line: the jump target has bit~0 cleared. In~(b), a sentence on the reserved bits of \texttt{rs2} is worded as a rule but does not name its instruction. The LLM supplies \texttt{HFENCE.GVMA} from the section it appears in, and the rule becomes self-contained (this is the rule behind \autoref{fig:hfence}). In~(c), the encoding of \texttt{HFENCE.GVMA} is a row of a table, and the rule lists its fixed bits and operand fields. Each of the three rules surfaced a bug in Table~\ref{tab:bugs} of Appendix~\ref{app:bugs}.
    
\subsection{Implementation Localization}
\label{sec:localize}

The code that implements one rule is only a small part in a large CPU repository. \tool{} extracts the corresponding code snippet as the \emph{code context} of the rule in two steps. 

\begin{figure}[t]
\centering
\includegraphics[width=\columnwidth]{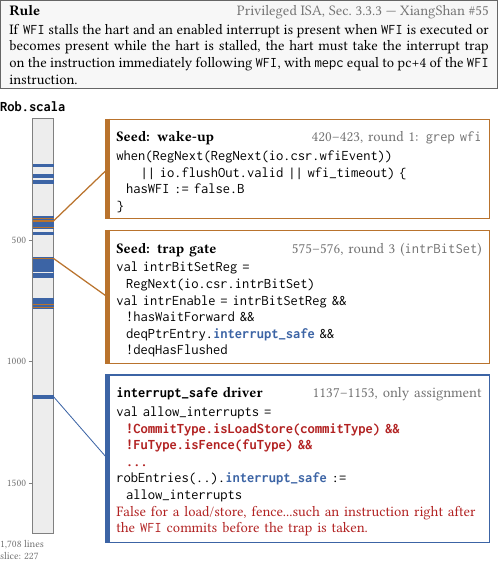}
\caption{Implementation localization for the rule behind bug~55 in Table~\ref{tab:bugs}: the seeds (orange), the other blocks the file slice kept (blue), and among them the code that decides which instructions are interruptible (red). The strip on the left is the file drawn to scale.}
\label{fig:localize}
\end{figure}

First, it understands the rule and searches from the repository for the core code snippets that are directly corresponding to the rule description. These snippets are called the \emph{seeds}. The seeds usually explicitly implement what the rule specifies, perform condition checks the rule requires, or assign the register or the field the rule concerns. 
From the rule, the LLM generates the patterns for a grep-based search and judges whether each hit is actually correlated to the rule and should be kept as a seed. 
However, searching for the seeds may not be completed within one round: the specification and the code sometimes name the same entity differently, a rule may involve several entities, and a seed may have to be traced back to the signals it depends on. We therefore design a ReAct-style agent~\cite{react}. 
In each round, the LLM builds the query based on what it collects and searches the code repository, until it considers that the seeds have covered the key implementation of the rule. 
For example, the rule behind bug~55 in Table~\ref{tab:bugs} requires that when an interrupt wakes a hart from \texttt{WFI}, the hart takes the interrupt before it executes the next instruction. The first round searched for the words of the rule: \texttt{InterruptEvent} returned nothing, and \texttt{wfi} returned 196 hits in 41 files, four of them related to the rule. From these the LLM learned the names the design uses, \texttt{hasWFI} for the sleeping hart and \texttt{wfiEvent} for the wake-up signal from the CSR unit, and the second round searched for them; its results showed a second signal from the CSR unit to the ROB, \texttt{intrBitSet}, the interrupt request. The third round searched for it and found the seed that decides the case: the ROB takes an interrupt only through \texttt{intrEnable}, which requires the instruction at its head to be interruptible (\autoref{fig:localize}, orange). The initial searches using terms from the rule did not retrieve these lines.

Second, the seeds are taken as the slicing criteria for extracting the code context. 
In practice, a seed by itself rarely suffices for the decision. Deciding whether the seed implements the rule correctly also requires the code related to it: where the signals it uses are set, where the value it produces is used, and the conditions under which it takes effect. \tool{} slices the seed's file to collect such code.
A file may run to thousands of lines, most of them about other parts of the design, and the relevant lines are easily lost in such an input~\cite{lostinmiddle}. \tool{} therefore uses LLM-based slicing~\cite{llmslicing,slicemate}, with one model call per file that holds a seed. 
The model reads the whole file, the rule and the seeds in the file, and is asked to extract only the code related to the seeds, filter out the unrelated code and return the blocks needed to make a decision.

In the example, the seeds show that waking up and taking the interrupt are separate paths: the wake-up only lets the hart resume, and the interrupt waits for \texttt{intrEnable}, which requires an interruptible head instruction. Which instructions are interruptible, the seeds do not say. The file slice kept \locsliceblocks{} blocks, \locslicelines{} lines, of the \locfilelines-line \texttt{Rob.scala} (blue), among them the code that decides it (red): loads, stores, fences, CSR accesses and atomics are never interruptible, since, as its comment explains, a memory access may already have reached a device. If a load follows the \texttt{WFI}, the woken hart therefore completes the load before it takes the interrupt: an instruction that should wait for the handler has executed, and the rule is violated.

\subsection{Candidate Identification}
\label{sec:identify}

The code context of a rule is confined to the files that hold its seeds, but whether the rule holds sometimes depends on code elsewhere in the repository. \tool{} therefore makes candidate identification retrieval-driven. Given the rule and the code context, the LLM may return one of three decisions: \emph{violation}, \emph{no violation}, or \emph{not enough code}. A decision of not enough code must list the questions blocking the decision, each asking for code that is absent from the context and can be found by searching the repository.

\begin{figure}[h]
\centering
\includegraphics[width=\columnwidth]{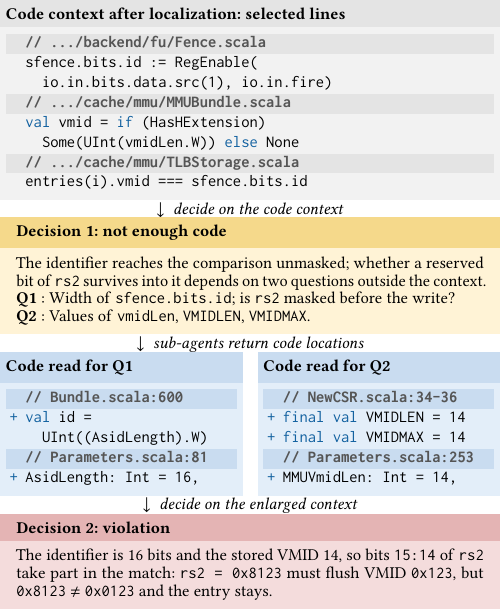}
\caption{Candidate identification for the rule of \autoref{fig:hfence},
in one run of \tool{}: its code context after localization (top), the
first decision with its two questions (Q1, Q2), one retrieval sub-agent
per question, the lines read back from the repository at the locations
they returned (marked~\texttt{+}), and the second decision.}
\label{fig:identify}
\end{figure}
    
Each question is assigned to its own ReAct-style retrieval sub-agent, so that each search is responsible for one question. The sub-agent searches the repository with \texttt{grep}, \texttt{glob}, and \texttt{read} until it judges that it has found the answer, then summarizes from what it has read the code facts that answer the question. To avoid the errors an LLM introduces when it copies code, the sub-agent reports each fact as the location of the code (a file and a line range), and \tool{} reads the lines at that location from the repository into the context. The LLM then decides again on the enlarged context. This repeats until the LLM decides violation or no violation, or the number of retrieval rounds reaches its limit. A decision of no violation ends the check of the rule. A decision of violation must explain how the code violates the rule, citing the lines involved; the rule, this analysis, and the code context form a \emph{candidate violation}, which enters the violation audit.

\autoref{fig:identify} shows this stage on the rule of \autoref{fig:hfence}. After localization, the code context held the path from \texttt{rs2} to the fence identifier and on to the VMID comparison in the TLB and page-table cache. Nothing on that path masks the identifier, but its width is declared in a file outside the context; had it been as wide as the VMID, the reserved bits would have been discarded and the rule satisfied. The first decision was therefore not enough code, with two questions: the width of the identifier and whether \texttt{rs2} is masked before the write (Q1), and the width of the stored VMID and the values of VMIDLEN and VMIDMAX named in the rule (Q2). The two sub-agents located the answers in the bundle declaration, the parameter file, and the CSR constants: the identifier is 16 bits wide, the VMID 14. On the enlarged context, the LLM decided violation and correctly identified its root cause.

\subsection{Violation Audit}
\label{sec:audit}

To keep the reported violations precise and trustworthy, \tool{} re-examines every candidate violation before reporting it. A candidate rests on two judgments by the LLM, its understanding of the rule and its understanding of the code, and each is open to a characteristic error. On the rule side, a rule read out of the context of the full specifications may be understood only partially: a provision elsewhere may state an exception for the very behavior the candidate concerns, and the LLM may also have misread the rule itself. On the code side, the LLM may have misread the code, or the situation the analysis describes may never arise in this design. The violation audit therefore checks the two judgments in two reviews: a \emph{specification-side review} over all candidates, followed by a \emph{code-side review} over the candidates it upholds.

Both reviews receive the rule, the code context, and the analysis from candidate identification of how the RTL violates the rule. Like seed search and the retrieval sub-agents, each review is a ReAct-style agent~\cite{react} with tools such as \texttt{grep} and \texttt{read} over its own source, the full specification corpus for one and the full repository for the other, and it searches that source to determine whether the reported violation holds.

\paragraph{Specification-side review.}
The LLM places the candidate back in the context of the full specifications, forms a complete understanding of what they require of the behavior in question, and decides whether that behavior still constitutes a violation. The LLM typically searches the specification corpus on its own initiative, using the rule's own text and the registers, fields, instructions, and modes named in the analysis as leads, and reads the passages they lead to; such a passage may state an exception for the behavior, or show that the specifications leave it to the implementation. A candidate the specifications uphold passes to the code-side review.

For example, the Privileged ISA~\cite{riscvpriv} states that a trap taken in M-mode while \texttt{mnstatus.NMIE} is zero puts the hart into a critical-error state, in which it stops and asserts a critical-error signal to the platform. Identification observed that XiangShan asserts this signal only while \texttt{dcsr.cetrig} is zero: with \texttt{cetrig} set, the hart enters the critical-error state but the signal stays low, and identification reported a violation. Searching the corpus for \texttt{cetrig}, the review found the deciding passage in the Debug Specification~\cite{riscvdebug}, in the field table of \texttt{dcsr}: when \texttt{cetrig} is set, a hart in a critical-error state enters Debug Mode instead of asserting the signal. The Privileged ISA refers to this alternative in its chapter on the Smdbltrp extension and leaves its definition to the Debug Specification. XiangShan enters Debug Mode in this case, and the review rejected the candidate.

\paragraph{Code-side review.}
This review takes the rule as given and independently verifies the analysis behind the candidate: whether the reported violation actually occurs in this core, whether it is observable, and whether the combination of extensions and configuration it relies on exists in this design. The review checks the analysis's reading of the cited code, verifies that the situation the analysis describes can arise, and confirms whether the deviation reaches architecturally observable state, a register, memory, a trap, or an output of the core. A candidate whose violation verdict stands is reported as a violation.

For example, a cache-block prefetch instruction is a hint: the CMO extension requires that it never raise an exception, and in particular that address translation for it neither check nor set the accessed and dirty bits of the page-table entry. Identification observed that the TLB of XiangShan applies its ordinary permission check to a prefetch and flags a page fault when the entry lacks the accessed or dirty bit, and it reported a violation. The review followed the flagged fault through the load pipeline: the pipeline clears every exception of a prefetch before the instruction commits, so the fault never becomes a trap. The review rejected the candidate.
         
\section{Evaluation}
\label{sec:eval}

Our evaluation answers four questions:
\begin{itemize}
\item \textbf{RQ1 (Effectiveness).} Does \tool{} find real,
previously unknown specification violations in real RISC-V CPUs, and what fraction of the candidates that survive its audit are true violations?
\item \textbf{RQ2 (Comparison with fuzzing).} Does CPU fuzzing find the bugs that \tool{} finds?
\item \textbf{RQ3 (Component contribution).} How much do the violation audit and rule-level checking contribute to the results?
\item \textbf{RQ4 (Cost).} What does a campaign cost in model usage
and time?
\end{itemize}

\subsection{Experimental Setup}
\label{sec:eval-setup}

We run \tool{} on two open-source RISC-V CPUs against four specification documents. We call one run of \tool{} on one CPU against one document a \emph{campaign}; there are six campaigns in total.

\paragraph{Audited CPUs.}
Table~\ref{tab:cores} lists the two CPUs. CVA6~\cite{cva6} is an application-class in-order CPU written in SystemVerilog; XiangShan~\cite{xiangshan} is a high-performance out-of-order CPU written in Chisel. Both are actively maintained, and they differ in HDL and microarchitecture, so together they test \tool{} on different kinds of RTL. We audit the source of each CPU at the commit listed in the table.

\begin{table}[tb]
\centering
\caption{Audited CPUs. \emph{Lines of RTL} counts the HDL sources in the tree given to \textbf{\textsc{\tool{}}}: \texttt{core/} of CVA6 and \texttt{src/main/scala/} of XiangShan (Kunminghu~V2).}
\label{tab:cores}
\small
\begin{tabular}{@{}l l l@{}}
\toprule
& \textbf{CVA6} & \textbf{XiangShan} \\
\midrule
HDL & SystemVerilog & Chisel \\
Microarchitecture & in-order & out-of-order \\
Audited commit & \texttt{b4d678f12} & \texttt{2e148b41d} \\
Commit date & 2026-01-09 & 2026-03-16 \\
Lines of RTL & $\sim$91{,}000 & $\sim$122{,}000 \\
\bottomrule
\end{tabular}
\end{table}

\paragraph{Specification corpus.}
We audit four documents published by RISC-V International: Volume~I (Unprivileged Architecture) and Volume~II (Privileged Architecture) of the RISC-V Instruction Set Manual, the RISC-V Advanced Interrupt Architecture (AIA), and the RISC-V Debug Specification. Table~\ref{tab:specs} gives the size of each document, the rules \tool{} extracts from it, and how many of them are in scope for each CPU.

\begin{table}[tb]
\centering
\caption{Specification documents, extracted rules, and rules in scope for each CPU. \emph{AIA}: Advanced Interrupt Architecture; \emph{Debug}: Debug Specification; \emph{XS}: XiangShan.}
\label{tab:specs}
\small
\setlength{\tabcolsep}{3pt}
\begin{tabular}{@{}l r r r r r@{}}
\toprule
& & & & \multicolumn{2}{c}{\textbf{In scope}} \\
\cmidrule(l){5-6}
\textbf{Document} & \textbf{Words} & \textbf{Sec.} &
\textbf{Rules} & \textbf{CVA6} & \textbf{XS} \\
\midrule
Unprivileged ISA & 146{,}476 & 664 & 2{,}715 & 1{,}212 & 1{,}310 \\
Privileged ISA & 79{,}789 & 244 & 1{,}758 & 1{,}188 & 1{,}431 \\
AIA & 37{,}998 & 103 & 540 & -- & 187 \\
Debug & 32{,}013 & 131 & 785 & -- & 357 \\
\midrule
Total & 296{,}276 & 1{,}142 & 5{,}798 & 2{,}400 & 3{,}285 \\
\bottomrule
\end{tabular}
\end{table}

\paragraph{Audit scope.}
A CPU cannot violate a rule about a feature it does not implement. For each CPU, we therefore determine the ISA extensions it implements from its documentation and configuration files, and audit only the sections that define these extensions (Appendix~\ref{app:scope}). The scope contains 410 sections with 2{,}400 rules for CVA6 and 580 sections with 3{,}285 rules for XiangShan (Table~\ref{tab:specs}).

\paragraph{Implementation.}
\tool{} is implemented in Python with LangGraph~\cite{langgraph}. Rules are checked in parallel, and so are the candidates within each review. Model calls use DeepSeek-V4-Flash~\cite{deepseekv4} (0731 release) in thinking mode with default sampling parameters; the reasoning effort is \emph{max}, except for the file slice, which uses \emph{high}. The documents are converted from PDF to Markdown with GLM-OCR~\cite{glmocr}.

\paragraph{Manual inspection and bug reports.}
We inspect every candidate that survives the audit and label it a \emph{true violation} or a \emph{false positive}. A candidate is a true violation if the RTL contradicts the rule and the contradicting behavior can occur in a project-supported configuration. (Two authors label each candidate independently and resolve disagreements by discussion.) Candidates that share one root cause are merged into one \emph{bug}, and we file one report per bug with the developers.

\subsection{RQ1: Effectiveness}
\label{sec:eval-bugs}

Table~\ref{tab:precision} summarizes the six campaigns. \tool{} checked \nrules{} rules in total. Identification produced \ncand{} candidate violations; \naudited{} of them survived the audit. Manual inspection labeled \ntrue{} of the \naudited{} audited candidates as true violations, a precision of \nprecall\%.

\begin{table*}[t]
\centering
\caption{Results of the six campaigns. \emph{Cand.}: candidate violations from identification. \emph{Audited}: candidates that survived the audit. \emph{True}: audited candidates labeled true violations by manual inspection. \emph{Prec.}: True/Audited. \emph{Bugs}: distinct bugs after merging candidates with one root cause, split into \emph{New} (first reported by us) and \emph{Dup} (reported earlier by others, or already fixed upstream). A bug found by two campaigns appears in both rows and is counted once in the total.}
\label{tab:precision}
\small
\setlength{\tabcolsep}{5pt}
\begin{tabular}{@{}l r r r r r r r r@{}}
\toprule
& & & & & & \multicolumn{3}{c}{\textbf{Bugs}} \\
\cmidrule(l){7-9}
\textbf{Campaign} & \textbf{Rules} & \textbf{Cand.} & \textbf{Audited} &
\textbf{True} & \textbf{Prec.\ (\%)} & \textbf{All} & \textbf{New} & \textbf{Dup} \\
\midrule
CVA6, Unprivileged ISA & 1{,}212 & 215 & 48 & 36 & 75.0 & 16 & 12 & 4 \\
CVA6, Privileged ISA & 1{,}188 & 289 & 104 & 76 & 73.1 & 35 & 15 & 20 \\
\midrule
XiangShan, Unprivileged ISA & 1{,}310 & 136 & 10 & 4 & 40.0 & 3 & 1 & 2 \\
XiangShan, Privileged ISA & 1{,}431 & 300 & 38 & 19 & 50.0 & 14 & 9 & 5 \\
XiangShan, Adv.\ Interrupt Arch. & 187 & 48 & 3 & 3 & 100.0 & 3 & 2 & 1 \\
XiangShan, Debug Specification & 357 & 95 & 14 & 10 & 71.4 & 7 & 6 & 1 \\
\midrule
Total & \nrules{} & \ncand{} & \naudited{} & \ntrue{} & \nprecall{} & \nbugsall{} & \nbugs{} & \ndup{} \\
\bottomrule
\end{tabular}
\end{table*}

\paragraph{From true violations to bugs.}
The \ntrue{} true violations correspond to \nbugsall{} distinct bugs, listed in Table~\ref{tab:bugs}, Appendix~\ref{app:bugs}.

\paragraph{Kinds of new bugs.}
The \nbugs{} new bugs come from all four documents. In six, the CPU accepts an encoding that must raise an illegal-instruction exception, because the encoding is reserved or defined only for the other base ISA width. In six, a legal instruction produces a wrong result or state update. In eight, a CSR field has a wrong reset value, a wrong set of legal values, or a wrong update on a write or a trap return; for example, \texttt{MRET} in CVA6 restores the virtualization mode \texttt{V} from \texttt{mstatus.MPV} even when it returns to M-mode, where \texttt{V} must be 0. In eight, the CPU takes the wrong trap, takes it late, or loses an interrupt. Seven are in address translation, permission checks, and fences, and seven are in the debug extension, from trigger matching to the halt sequence out of reset.

\paragraph{False positives.}
Of the 69 audited candidates that manual inspection rejected, 24 read the specification too strictly: the rule has a condition or a more permissive reading elsewhere in the corpus, such as the permission to implement \texttt{time} by trapping and emulating. Thirty are wrong about the RTL: the candidate misreads the code it cites, or the deviation it describes cannot arise or be observed in this core; one, for instance, reports that CVA6 reads the CSR on a \texttt{CSRRW} with \texttt{rd}=\texttt{x0}, which the ISA forbids, but the read has no side effect and its result is discarded, so software cannot tell. Fifteen hold for the source but do not occur in practice: 14 depend on parameter combinations unsupported by the audited implementation, such as RV32 with the hypervisor extension in CVA6, and one on a state the software stack never produces (an execute trigger armed while the hart single-steps, which the standard debugger rules out). The audit removes \nrejected{} of the \ncand{} candidates (Section~\ref{sec:eval-ablation}); in the 69 false positives it keeps, the specification-side and the code-side review read the specification or the RTL the same way the detector did.

\textbf{Answer to RQ1.} \tool{} found \nbugsall{} bugs in two RISC-V CPUs, \nbugs{} previously unknown, and \nprecall\% of the candidates that survive its audit are true violations.

\subsection{RQ2: Comparison with CPU Fuzzing}
\label{sec:eval-fuzz}

\paragraph{Protocol.}
We ran DiveFuzz with its default generation configuration for \nfuzzh{} hours per CPU on the commits of Table~\ref{tab:cores}, using the co-simulation flow of each project to execute the programs and compare them against the reference model (Spike for CVA6, NEMU for XiangShan). The runs produced 320{,}384 programs for CVA6 and 41{,}984 for XiangShan. We examined every divergence, reading the two traces against the RTL, the reference model, and the specification, and matched the RTL bugs among them against Table~\ref{tab:bugs}.

\begin{table}[tb]
\centering
\caption{Bugs of Table~\ref{tab:bugs} that \nfuzzh{} hours of DiveFuzz found on the audited commits, among the \nbugs{} new bugs (\emph{New}) and all \nbugsall{} bugs (\emph{All}).}
\label{tab:fuzzbaseline}
\small
\setlength{\tabcolsep}{5pt}
\begin{tabular}{@{}l r r r@{}}
\toprule
\textbf{CPU} & \textbf{Programs} & \textbf{New} & \textbf{All} \\
\midrule
CVA6 & 320{,}384 & 0 & 2 \\
XiangShan & 41{,}984 & 0 & 0 \\
\midrule
Total & 362{,}368 & 0 / \nbugs{} & 2 / \nbugsall{} \\
\bottomrule
\end{tabular}
\end{table}

\paragraph{Results.}
DiveFuzz found two of the \nbugsall{} bugs and none of the \nbugs{} new ones (Table~\ref{tab:fuzzbaseline}). \texttt{ecall} and \texttt{ebreak} write the instruction word into \texttt{mtval} (row~29), and the big-endian control bits of \texttt{mstatus} and \texttt{hstatus} are writable although the CPU has no big-endian data path (row~19). DiveFuzz also found four bugs that \tool{} did not report, all of them already reported upstream.

\textbf{Answer to RQ2.} DiveFuzz found none of \tool{}'s \nbugs{} new bugs in \nfuzzh{} hours per CPU.

\subsection{RQ3: Component Contribution}
\label{sec:eval-ablation}

RQ3 measures the contribution of two design decisions: the violation audit (Section~\ref{sec:audit}) and checking one rule at a time instead of a whole section (Section~\ref{sec:extract}).

\paragraph{Violation audit.}
We ran each review independently on all \ncand{} candidates and compared four configurations: no audit, the specification-side review alone, the code-side review alone, and the two reviews in sequence, which is the pipeline. Table~\ref{tab:audit-ablation} reports, for each configuration, the number of candidates it keeps and the precision among them. Only the candidates that the pipeline keeps were inspected (Section~\ref{sec:eval-setup}), so the precision of the other three configurations is a lower bound.

\begin{table}[t]
\centering
\caption{Contribution of the two reviews. For each audit configuration, the candidates it keeps (\emph{Kept}) and the share of true violations among them (\emph{Prec.}, \%; see text). \emph{Both} is the pipeline. \emph{Unpriv}, \emph{Priv}, \emph{AIA}, \emph{Debug}: the four documents; \emph{XS}: XiangShan.}
\label{tab:audit-ablation}
\small
\setlength{\tabcolsep}{3.2pt}
\begin{tabular}{@{}l r r r r r r r r@{}}
\toprule
& \multicolumn{2}{c}{\textbf{No audit}} &
\multicolumn{2}{c}{\textbf{Spec.\ only}} &
\multicolumn{2}{c}{\textbf{Code only}} &
\multicolumn{2}{c}{\textbf{Both}} \\
\cmidrule(lr){2-3} \cmidrule(lr){4-5} \cmidrule(lr){6-7} \cmidrule(l){8-9}
\textbf{Campaign} & Kept & Prec. & Kept & Prec. & Kept & Prec. & Kept & Prec. \\
\midrule
CVA6 Unpriv & 215 & 16.7 & 92 & 39.1 & 101 & 35.6 & 48 & 75.0 \\
CVA6 Priv & 289 & 26.3 & 142 & 53.5 & 179 & 42.5 & 104 & 73.1 \\
XS Unpriv & 136 & 2.9 & 29 & 13.8 & 39 & 10.3 & 10 & 40.0 \\
XS Priv & 300 & 6.3 & 79 & 24.1 & 136 & 14.0 & 38 & 50.0 \\
XS AIA & 48 & 6.3 & 14 & 21.4 & 12 & 25.0 & 3 & 100.0 \\
XS Debug & 95 & 10.5 & 28 & 35.7 & 33 & 30.3 & 14 & 71.4 \\
\midrule
Total & \ncand{} & \nprecnone{} & \nspeckept{} & \nprecspec{} & \ncodekept{} & \nprecode{} & \naudited{} & \nprecall{} \\
\bottomrule
\end{tabular}
\end{table}

\begin{table*}[t]
\centering
\caption{Whole-section against rule-level checking, on the 94 sections that contain a rule behind a bug of Table~\ref{tab:precision}. \emph{Sections}, \emph{Rules}: the sections given to the variant and the rules they contain. \emph{Bugs}: bugs that rule-level checking found in these sections; a bug found by two campaigns is counted in one row. \emph{Cand.}: candidates the variant produced. \emph{Audited}: candidates that survived the audit, split by manual inspection into true violations (\emph{True}) and false positives (\emph{False}). \emph{Found}: bugs covered by an audited candidate labeled a true violation. \emph{Recall}: Found over Bugs.}
\label{tab:wholesec}
\small
\setlength{\tabcolsep}{4pt}
\begin{tabular}{@{}l r r r r r r r r r@{}}
\toprule
\textbf{Campaign} & \textbf{Sections} & \textbf{Rules} & \textbf{Bugs} & \textbf{Cand.} & \textbf{Audited} & \textbf{True} & \textbf{False} & \textbf{Found} & \textbf{Recall (\%)} \\
\midrule
CVA6, Unprivileged ISA & 24 & 184 & 15 & 19 & 18 & 16 & 2 & 9 & 60.0 \\
CVA6, Privileged ISA & 41 & 626 & 33 & 32 & 25 & 23 & 2 & 16 & 48.5 \\
\midrule
XiangShan, Unprivileged ISA & 4 & 24 & 3 & 3 & 2 & 2 & 0 & 1 & 33.3 \\
XiangShan, Privileged ISA & 16 & 178 & 12 & 10 & 8 & 7 & 1 & 4 & 33.3 \\
XiangShan, Adv.\ Interrupt Arch. & 3 & 16 & 3 & 3 & 2 & 2 & 0 & 2 & 66.7 \\
XiangShan, Debug Specification & 6 & 110 & 7 & 4 & 2 & 2 & 0 & 2 & 28.6 \\
\midrule
Total & 94 & 1{,}138 & \nbugsall{} & 71 & 57 & 52 & 5 & 34 & 46.6 \\
\bottomrule
\end{tabular}
\end{table*}

The two reviews reject largely different candidates: of the \nrejected{} candidates that the pipeline rejects, only 416 are rejected by both reviews. Either review alone would leave \nspeckept{} or \ncodekept{} candidates to inspect, at a precision of \nprecspec\% or \nprecode\%; the two in sequence leave \naudited{} candidates at \nprecall\%, compared with \nprecnone\% without the audit.

\paragraph{Rule-level checking.}
The whole-section variant treats an entire specification section as one rule: localization, identification, and the audit run unchanged, with the same model, prompts, and tools, but on the section instead of one of its rules. This variant corresponds to applying the LLM to the specification text directly. We ran it on the 94 sections that contain a rule behind one of the \nbugsall{} bugs of Section~\ref{sec:eval-bugs}; these sections contain 1{,}138 rules. Because the sections were selected from the bugs that rule-level checking found, rule-level checking finds all \nbugsall{} bugs on them, and Table~\ref{tab:wholesec} reports how many of them the variant finds.

The variant produced 71 candidates, of which 57 survived the audit. Of these 57, 52 are true violations, and they cover 34 of the \nbugsall{} bugs, a recall of 46.6\%.

\textbf{Answer to RQ3.} The audit reduces the candidates to inspect from \ncand{} to \naudited{} and raises precision from \nprecnone\% to \nprecall\%. Checking one rule at a time finds 39 bugs that whole-section checking misses.

\begin{table*}[t]
\centering
\caption{Cost of the six campaigns: \emph{Localization + identification} over all rules of a campaign, \emph{Audit} over both reviews. Tokens in millions, input~\textbar~output, output including reasoning. \emph{Per rule}: mean time from seed search to final decision. Rule extraction (\$7.9 in total) is excluded.}
\label{tab:cost}
\small
\setlength{\tabcolsep}{4pt}
\begin{tabular}{@{}l r r P r r r r P r r@{}}
\toprule
& & \multicolumn{5}{c}{\textbf{Localization + identification}} &
\multicolumn{5}{c}{\textbf{Audit}} & \\
\cmidrule(lr){3-7} \cmidrule(lr){8-12}
\textbf{Campaign} & \textbf{Rules} & Calls &
\multicolumn{2}{c}{\hd{Tokens (M)}{\inout}} & \hd{Per rule}{(min)} &
\hd{Cost}{(\$)} & Cand. & Calls &
\multicolumn{2}{c}{\hd{Tokens (M)}{\inout}} & \hd{Cost}{(\$)} &
\hd{\textbf{Total}}{\textbf{(\$)}} \\
\midrule
CVA6, Unprivileged ISA & 1{,}212 & 11{,}808 & 132.8 & 263.3 & 20.5 & 186.5 & 215 & 2{,}971 & 47.0 & 10.6 & 8.4 & 194.9 \\
CVA6, Privileged ISA & 1{,}188 & 11{,}378 & 141.8 & 225.0 & 19.2 & 160.5 & 289 & 4{,}439 & 71.8 & 15.3 & 12.6 & 173.1 \\
\midrule
XiangShan, Unprivileged ISA & 1{,}310 & 18{,}806 & 191.1 & 308.9 & 23.2 & 223.0 & 136 & 1{,}426 & 24.8 & 4.7 & 4.0 & 227.0 \\
XiangShan, Privileged ISA & 1{,}431 & 18{,}315 & 185.7 & 304.8 & 19.6 & 220.3 & 300 & 3{,}506 & 57.8 & 10.5 & 9.0 & 229.3 \\
XiangShan, Adv.\ Interrupt Arch. & 187 & 2{,}153 & 20.0 & 37.9 & 20.3 & 27.1 & 48 & 611 & 11.2 & 2.4 & 1.9 & 29.0 \\
XiangShan, Debug Specification & 357 & 3{,}988 & 36.9 & 73.4 & 19.1 & 52.6 & 95 & 1{,}267 & 21.8 & 3.7 & 3.2 & 55.7 \\
\midrule
Total & \nrules{} & 66{,}448 & 708.3 & 1{,}213.2 & 20.5 & 870.0 & \ncand{} & 14{,}220 & 234.2 & 47.0 & 39.0 & 909.1 \\
\bottomrule
\end{tabular}
\end{table*}

\subsection{RQ4: Cost}
\label{sec:eval-cost}

Table~\ref{tab:cost} reports the model usage, time, and cost of each campaign; Appendix~\ref{app:cost} further breaks these figures down by step. Costs are computed from the provider's list prices at the time of the runs (all runs fell in the provider's off-peak hours): \$0.007 per million input tokens served from the prompt cache, \$0.22 per million other input tokens, and \$0.66 per million output tokens, which include the model's reasoning tokens.

\paragraph{Cost and time per campaign.}
The six campaigns cost \$909 in total, or about \$0.16 per checked rule. Individual campaigns cost between \$29 for XiangShan AIA and \$229 for XiangShan Privileged ISA, largely reflecting the number and complexity of rules in each specification. Across all campaigns, \tool{} makes 66,448 model calls for localization and identification, processing 708.3 million input tokens and 1,213.2 million output tokens. A rule takes 20.5 minutes on average from seed search to its final decision, with the per-campaign average ranging from 19.1 to 23.2 minutes.
Most of the cost comes from localization and identification, which together account for \$870, or 95.7\% of the total cost. In comparison, the violation audit processes 1,083 candidate violations with 14,220 additional model calls and costs only \$39 in total, about 4.3\% of the overall budget. Thus, although the audit substantially filters candidate violations before manual inspection, it adds relatively little monetary cost compared with locating and reasoning about the RTL implementation of each specification rule.

\textbf{Answer to RQ4.} One campaign costs at most \$230, and a rule takes about 20 minutes. Localization and identification dominate the cost; the audit adds 4\%.

\subsection{Limitations}
\label{sec:limitations}

Our results are subject to the following limitations. We can measure
precision but not recall: the number of specification violations in
either CPU is unknown, and RQ3 measures the rule-level design only
against the bugs the pipeline itself found. Identification and both
reviews run on the same model, so a reading the detector gets wrong can
be repeated by the reviewer.
     
\section{Related Work}
\label{sec:related}

\paragraph{CPU fuzzing.}
CPU fuzzers run generated programs on the RTL and a reference model and report architectural divergences (Section~\ref{sec:motivate-existing}). They differ in test generation and guidance: coverage feedback~\cite{difuzzrtl,thehuzz,processorfuzz}, runtime instruction rewriting~\cite{morfuzz}, richer instruction mixes and dependences~\cite{cascade,divefuzz,riscsmith}, data-sensitive operands and state-transition feedback~\cite{drvfuzz}, or generative models trained on hardware feedback~\cite{genhuzz}. Nevertheless, detection requires both a program that triggers the bug and a reference model that implements the relevant behavior correctly. Bugs may be missed because of limited input generation or incomplete reference-model behavior. Encarsia~\cite{encarsia} further shows that current fuzzers miss many injected bugs even when they are visible in architectural state. In contrast, \tool{} derives correctness directly from the RISC-V specifications, requiring neither a reference model nor a triggering program.

\paragraph{Formal verification and static analysis.}
Formal verification proves that a CPU satisfies a property written for the checker, whether on single instructions~\cite{riscvformal} or on conformance to an ISA-level model~\cite{isaformal}. Static analysis tools can check RTL against predefined bug patterns, such as CWE weaknesses~\cite{cweat}, missing resets, unreachable states of a finite-state machine, or a bit-width mismatch that truncates a value~\cite{qihe,cweat}; ChiSA~\cite{chisa} checks the assertions that developers write into Chisel code. These generic patterns often do not encode specification-specific semantics, limiting their ability to directly detect specification violations beyond their predefined scope. These approaches rely on predefined properties or bug patterns; \tool{} checks RTL directly against the RISC-V specifications without requiring such separately constructed artifacts.

\paragraph{LLMs for hardware design and verification.}
LLMs write Verilog from natural-language descriptions~\cite{verigen} and fix its syntax errors from compiler feedback~\cite{rtlfixer}. In verification, AssertLLM~\cite{assertllm} turns a design document into SystemVerilog assertions for a model checker, UVLLM~\cite{uvllm} generates UVM tests and repairs the design from simulation failures, FLAG~\cite{flag} flags likely-buggy lines in RTL without running a test, and VerilogLAVD~\cite{veriloglavd} uses an LLM to write bug detection patterns that a fixed engine runs over Verilog. In these works the LLM starts from a design document, the CWE list, or the code of one module. \tool{} starts from the RISC-V specifications and CPU repository: the LLM reads a rule and the code that implements it, built up step by step from the repository, and decides whether the code violates the rule.
        
\section{Discussion}
\label{sec:discussion}

\textbf{Limitations of reference-model-based differential testing.}
XiangShan uses NEMU~\cite{nemu} as its reference model. NEMU is the RISC-V ISA simulator maintained by the XiangShan team; DiffTest~\cite{difftest}, the differential testing framework of XiangShan, compares the architectural state of the RTL against it after every committed instruction, and DiveFuzz used it as the oracle for XiangShan in Section~\ref{sec:eval-fuzz}. We checked each of the 25 XiangShan bugs of Table~\ref{tab:bugs} (rows 49--73) against the NEMU source. The bugs fall into three groups. First, nine bugs were also present in NEMU when we identified the corresponding bugs in XiangShan: a misaligned access to a non-idempotent region raises address-misaligned instead of access-fault, \texttt{HLVX} checks only read permission at the PMP/PMA stage, and address triggers compare the address before pointer masking. In these cases, both the RTL and the reference model exhibited the same incorrect behavior, leaving no discrepancy for differential testing to detect. Second, seven bugs concern behavior that NEMU simplifies away: NEMU implements the Sdtrig triggers but has no Debug Module and no Debug Mode (\texttt{dmode} is hardwired to zero, and no trigger action can enter Debug Mode), its performance counters are read-only zero and \texttt{mip.LCOFIP} is copied from the RTL (\autoref{fig:lcofi}), an interrupt is injected only when the RTL commits one and with the number the RTL chose, and \texttt{WFI} is a no-op. These behaviors are never compared. Third, for the remaining nine bugs, NEMU follows the specification, so differential testing can detect them in principle, but triggering them still requires carefully constructed programs. By checking RTL directly against the RISC-V specifications, \tool{} identifies violations without relying on separately constructed reference models, formal properties, or bug patterns, and without requiring triggering programs.
     
\section{Conclusion}
\label{sec:concl}

This paper presents \tool{}, which checks CPU RTL directly against the RISC-V specifications without relying on reference models, manually written properties, or predefined bug patterns. \tool{} extracts specification rules, locates their RTL implementations, identifies potential violations with on-demand context retrieval, and audits candidates before reporting them. On CVA6 and XiangShan, \tool{} reported \naudited{} candidates with \nprecall\% precision, corresponding to \nbugsall{} distinct bugs, including \nbugs{} previously unknown bugs; developers have fixed \nfixed{} and confirmed \nconfirmed{} more.

\bibliographystyle{ACM-Reference-Format}
\bibliography{refs}

\clearpage
\appendix
\onecolumn
\section{List of Bugs}
\label{app:bugs}

Table~\ref{tab:bugs} lists the \nbugsall{} bugs summarized in
Section~\ref{sec:eval-bugs}. Each row is one distinct bug after
merging candidates that share a cause; the section column gives the
section of the rule that surfaced it, and when a bug surfaced through
rules in several sections we list the most specific one.

{\footnotesize
\setlength{\tabcolsep}{4pt}
\begin{xltabular}{\textwidth}{@{}r l l l X l l@{}}
\caption{Bugs found by \textbf{\textsc{\tool{}}} (\nbugsall{} bugs: \nbugs{} new,
\ndup{} duplicates). \emph{Spec.} and \emph{Section}: the specification
document and the section from which the violated rule was extracted;
\emph{Unpriv} and \emph{Priv} are Volumes~I and~II of the instruction set
manual, \emph{AIA} the Advanced Interrupt Architecture, \emph{Debug} the
Debug Specification. \emph{New}:
we were the first to report the bug; \emph{Dup}: it had been reported by
others before our manual review. Status as of 2026-09-09:
\emph{Fixed}, a fix has landed upstream; \emph{Confirmed}, maintainers
have accepted the report; \emph{Reported} otherwise. For new bugs,
\textbf{Fixed} and \textbf{Confirmed} are in bold.}
\label{tab:bugs}\\
\toprule
\textbf{\#} & \textbf{Core} & \textbf{Spec.} & \textbf{Section} &
\textbf{Description} & \textbf{New/Dup} & \textbf{Status} \\
\midrule
\endfirsthead
\caption[]{Bugs found by \textbf{\textsc{\tool{}}} (continued).}\\
\toprule
\textbf{\#} & \textbf{Core} & \textbf{Spec.} & \textbf{Section} &
\textbf{Description} & \textbf{New/Dup} & \textbf{Status} \\
\midrule
\endhead
\midrule
\multicolumn{7}{r}{\emph{Continued on next page}}\\
\endfoot
\bottomrule
\endlastfoot
1 & CVA6 & Unpriv & 7.2 & On RV32, a U-mode read of \texttt{hpmcounter3h} returns zero because the address check uses $>$ instead of $\geq$. & Dup & Reported \\
2 & CVA6 & Unpriv & 13.2 & SC in \texttt{WAIT\_AW\_READY} drops \texttt{aw.lock}; memory is written while \texttt{rd} reports failure. & \textbf{New} & \textbf{Fixed} \\
3 & CVA6 & Unpriv & 17.4.2.2 & CBO instructions take ordinary store permission checks instead of CBO checks. & \textbf{New} & Reported \\
4 & CVA6 & Unpriv & 17.5 & Under H, the CBO permission check ignores \texttt{V}, so illegal- and virtual-instruction are decided wrongly. & \textbf{New} & \textbf{Confirmed} \\
5 & CVA6 & Unpriv & 18.7 & RV32 with F/D accepts the RV64-only \texttt{FCVT.L}/\texttt{FCVT.LU} conversions instead of raising illegal-instruction. & \textbf{New} & \textbf{Fixed} \\
6 & CVA6 & Unpriv & 26.1 & With Zcmp and floating point both enabled, Zcmp encodings are decoded as \texttt{c.fsdsp}. & Dup & Fixed \\
7 & CVA6 & Unpriv & 26.13.3 & The Zcmp macro decoder is not flushed on a trap, so expansion resumes from a stale intermediate state. & Dup & Confirmed \\
8 & CVA6 & Unpriv & 26.13.6 & The Zcmp macro decoder leaves \texttt{IDLE} without waiting for \texttt{issue\_ack}, so the first expanded instruction can be lost. & \textbf{New} & \textbf{Confirmed} \\
9 & CVA6 & Unpriv & 26.13.7 & RV64 Zcmp push/pop steps \texttt{sp} by four bytes, overlapping stack slots. & \textbf{New} & Reported \\
10 & CVA6 & Unpriv & 26.13.9 & Reserved Zcmp \texttt{rlist} values 0--3 are accepted instead of raising illegal-instruction; the expansion underflows \texttt{sp} and can hang. & \textbf{New} & \textbf{Fixed} \\
11 & CVA6 & Unpriv & 26.14 & The JVT table index is truncated to six bits, so larger indices wrap. & \textbf{New} & \textbf{Fixed} \\
12 & CVA6 & Unpriv & 26.14.2 & The Zcmt JVT entry is fetched by a plain D-cache read, with no execute or PMA check and no way to raise a fetch fault. & \textbf{New} & Reported \\
13 & CVA6 & Unpriv & 26.14.5 & \texttt{cm.jt}/\texttt{cm.jalt} leave bit~0 of the jump target set. & \textbf{New} & \textbf{Fixed} \\
14 & CVA6 & Unpriv & 27.9.17 & Bit-manipulation encodings with reserved fields are accepted instead of raising illegal-instruction. & \textbf{New} & \textbf{Fixed} \\
15 & CVA6 & Unpriv & 27.9.46 & The decoder has no XLEN guard on Zbkb/Zknh encodings: RV64 accepts the RV32-only \texttt{zip}/\texttt{unzip} and SHA-512 instructions, and RV32 the RV64-only ones. & \textbf{New} & \textbf{Fixed} \\
16 & CVA6 & Priv & 2.2.3 & A VS-CSR write under \texttt{V}=1 raises virtual-instruction but still commits. & \textbf{New} & Reported \\
17 & CVA6 & Priv & 3.1.6.3 & \texttt{SXL}/\texttt{UXL} still read as 64 when S/U modes are not implemented. & \textbf{New} & \textbf{Fixed} \\
18 & CVA6 & Priv & 3.1.6.4 & Without H, \texttt{MRET}/\texttt{SRET} do not clear \texttt{mstatus.MPRV}. & \textbf{New} & Reported \\
19 & CVA6 & Priv & 3.1.6.5 & \texttt{mstatus.MBE}/\texttt{SBE} and \texttt{hstatus.VSBE} are writable, but no big-endian data path exists. & Dup & Reported \\
20 & CVA6 & Priv & 3.1.6.5 & Without S-mode, \texttt{mstatus.SBE} is writable and reads back nonzero instead of being read-only zero. & Dup & Reported \\
21 & CVA6 & Priv & 3.1.6.7 & \texttt{mstatus.SD} stays clear when vector state is dirty. & \textbf{New} & Reported \\
22 & CVA6 & Priv & 3.1.6.7 & \texttt{fcvt} does not set \texttt{mstatus.FS} to Dirty when it implicitly modifies \texttt{fcsr}. & Dup & Reported \\
23 & CVA6 & Priv & 3.1.9 & Instructions already decoded when a write to \texttt{mie} enables an interrupt retire before the interrupt is taken. & Dup & Reported \\
24 & CVA6 & Priv & 3.1.9 & The interrupt priority chain ignores \texttt{mideleg}, so a pending SEI masks a delegated supervisor timer interrupt. & Dup & Reported \\
25 & CVA6 & Priv & 3.1.11 & Under H, a VU-mode counter access with \texttt{mcounteren}=1, \texttt{hcounteren}=1 and \texttt{scounteren}=0 does not trap. & Dup & Reported \\
26 & CVA6 & Priv & 3.1.14 & Without C, writes to \texttt{sepc}/\texttt{mepc}/\texttt{vsepc} leave bit~1 set. & \textbf{New} & \textbf{Fixed} \\
27 & CVA6 & Priv & 3.1.15 & A misaligned \texttt{LR} raises store-address-misaligned instead of load-address-misaligned. & Dup & Reported \\
28 & CVA6 & Priv & 3.1.15 & When a page fault and a PMP denial coincide, an access-fault is reported instead of the page-fault. & Dup & Confirmed \\
29 & CVA6 & Priv & 3.1.16 & \texttt{ecall}/\texttt{ebreak} write the instruction word into \texttt{mtval}. & Dup & Confirmed \\
30 & CVA6 & Priv & 3.1.16 & A misaligned branch or jump target reports the branch \texttt{pc} in \texttt{tval} instead of the target address. & Dup & Fixed \\
31 & CVA6 & Priv & 3.1.18 & \texttt{menvcfg}/\texttt{senvcfg.CBIE} accept the reserved encoding \texttt{10} and read it back. & Dup & Fixed \\
32 & CVA6 & Priv & 3.6 & The load unit drops the page offset when classifying speculative loads to non-idempotent regions. & Dup & Reported \\
33 & CVA6 & Priv & 3.7.1.1 & TOR address matching uses \texttt{pmpaddr} bit~0, which lies below the PMP granularity \texttt{G} and must not affect the match. & Dup & Reported \\
34 & CVA6 & Priv & 3.7.1.3 & An 8-byte S-mode load/store that crosses a TOR PMP region boundary is accepted. & Dup & Fixed \\
35 & CVA6 & Priv & 4.2.1 & Writing \texttt{mstatus.MXR} does not invalidate DTLB entries that cached the old permission. & \textbf{New} & \textbf{Confirmed} \\
36 & CVA6 & Priv & 4.2.1 & VS-mode \texttt{SFENCE.VMA} with \texttt{rd}$\neq$0 retires instead of raising illegal-instruction. & Dup & Fixed \\
37 & CVA6 & Priv & 4.3.2 & A store-side PTW PMP failure is reported as a load access-fault. & \textbf{New} & Reported \\
38 & CVA6 & Priv & 4.3.2 & The G-stage leaf that maps a VS-stage page table truncates the host PPN to the guest PPN width. & \textbf{New} & \textbf{Confirmed} \\
39 & CVA6 & Priv & 5.1 & \texttt{MRET} with \texttt{MPP}=M restores \texttt{V} from \texttt{MPV} instead of leaving \texttt{V}=0. & \textbf{New} & \textbf{Fixed} \\
40 & CVA6 & Priv & 5.2.10 & A write to \texttt{hgatp} with an unsupported \texttt{MODE} is dropped entirely, so \texttt{PPN} and \texttt{VMID} keep their old values. & Dup & Reported \\
41 & CVA6 & Priv & 5.3.2 & \texttt{HFENCE} with \texttt{rd}$\neq$0 is executed instead of being treated as illegal. & \textbf{New} & \textbf{Fixed} \\
42 & CVA6 & Priv & 5.4.1 & \texttt{MPRV}=1 with \texttt{MPV}=1 and \texttt{MPP}=M still translates through VS/G-stage. & \textbf{New} & Reported \\
43 & CVA6 & Priv & 5.4.1 & The illegal-instruction trap raised on CV-X-IF rejection sets \texttt{GVA} from the virtualization mode. & Dup & Fixed \\
44 & CVA6 & Priv & 5.4.2 & Under H, the VS bits of \texttt{mideleg} are not read-only one after reset. & \textbf{New} & \textbf{Fixed} \\
45 & CVA6 & Priv & 5.4.4 & \texttt{mtval2} accepts and reads back arbitrary values, outside its WARL legal set. & \textbf{New} & \textbf{Confirmed} \\
46 & CVA6 & Priv & 5.5.1 & The G-stage check on a VS-stage page-table read does not test the \texttt{U} bit of the G-stage leaf PTE. & Dup & Reported \\
47 & CVA6 & Priv & 5.6.3 & \texttt{mtinst}/\texttt{htinst} are written with a transformed instruction on illegal-/virtual-instruction traps and on interrupts, where they must be zero. & Dup & Reported \\
48 & CVA6 & Priv & 11 & \texttt{HLV.D} is decoded without checking that its reserved \texttt{rs2} field is zero or that XLEN is 64, so non-conforming encodings retire instead of raising illegal-instruction. & \textbf{New} & Reported \\
\midrule
49 & XiangShan & Unpriv & 1.4 & A misaligned load/store whose second half crosses into the non-canonical address hole is translated and executed instead of faulting. & \textbf{New} & \textbf{Confirmed} \\
50 & XiangShan & Unpriv & 17.6.1 & A CBO to a page with \texttt{PBMT}=NC raises store access-fault instead of executing. & Dup & Fixed \\
51 & XiangShan & Unpriv & 19.2 & \texttt{FLH} through the uncached path does not NaN-box the loaded half-precision value. & Dup & Fixed \\
52 & XiangShan & Priv & 3.1.6.2 & \texttt{MRET} returning to VU does not clear \texttt{vsstatus.SDT}. & \textbf{New} & \textbf{Fixed} \\
53 & XiangShan & Priv & 3.1.8 & With \texttt{mnstatus.NMIE}=0, the HS trap entry is gated off, so an exception delegated to S/U-mode becomes a critical error instead of trapping to \texttt{stvec}. & \textbf{New} & Reported \\
54 & XiangShan & Priv & 3.1.10 & A write to \texttt{minstret} or another counter CSR takes effect at issue rather than after the writing instruction completes, so \texttt{csrw minstret, X} followed by \texttt{csrr} reads $X{+}1$. & \textbf{New} & Reported \\
55 & XiangShan & Priv & 3.3.3 & After \texttt{WFI} is woken by an interrupt, the next load/store/CSR/fence executes before the trap is taken, so \texttt{mepc} points past it. & \textbf{New} & \textbf{Confirmed} \\
56 & XiangShan & Priv & 3.4 & \texttt{mnstatus.NMIE} resets to 1 and can be written to 0, and \texttt{mstatus.MDT} resets to 0. & Dup & Fixed \\
57 & XiangShan & Priv & 3.6.7 & Misaligned accesses to non-idempotent regions raise address-misaligned instead of access-fault. & \textbf{New} & \textbf{Fixed} \\
58 & XiangShan & Priv & 4.5.1 & Sequential fetch that falls through across the Sv39/Sv48 canonical boundary does not raise an instruction page-fault. & Dup & Fixed \\
59 & XiangShan & Priv & 5.2 & A VS-mode access to \texttt{vstimecmp} with \texttt{menvcfg.STCE}=0 raises illegal-instruction where \texttt{henvcfg.STCE} requires virtual-instruction. & Dup & Fixed \\
60 & XiangShan & Priv & 5.3.1 & \texttt{HLVX} checks only read permission at the final PMP/PMA stage, not execute permission. & Dup & Fixed \\
61 & XiangShan & Priv & 5.3.2 & \texttt{HFENCE.GVMA} matches VMID against reserved \texttt{rs2} bits above VMIDMAX, so selective fences can miss. & \textbf{New} & \textbf{Fixed} \\
62 & XiangShan & Priv & 5.5.3 & \texttt{HFENCE.GVMA x0,x0} leaves VS-stage entries in the page-table cache, so they survive a G-stage change. & Dup & Fixed \\
63 & XiangShan & Priv & 6.10.2.6 & Address-trigger comparison uses the unmasked pointer, ignoring pointer masking. & \textbf{New} & \textbf{Fixed} \\
64 & XiangShan & Priv & 9.1.1 & A counter-overflow interrupt request that coincides with a CSR write to \texttt{mip}/\texttt{sip} is dropped, so \texttt{LCOFIP} is never set. & \textbf{New} & \textbf{Confirmed} \\
65 & XiangShan & AIA & 5.5 & \texttt{WFI} does not wake on interrupts pending in \texttt{hvictl}/\texttt{vstopi}. & \textbf{New} & \textbf{Confirmed} \\
66 & XiangShan & AIA & 6.3.3 & A dead enable term in the \texttt{vstopi} logic makes the priority-255 encoding for a lone VSEI candidate unreachable. & Dup & Fixed \\
67 & XiangShan & Debug & 4.6 & \texttt{haltreq}/\texttt{resethaltreq} out of reset has no dedicated halt path, so the hart may commit its first instruction before halting. & \textbf{New} & \textbf{Confirmed} \\
68 & XiangShan & Debug & 4.9.1 & \texttt{dcsr.stopcount}=1 does not stop \texttt{minstret}/\texttt{mcycle}; only the \texttt{instret}/\texttt{cycle} read values are frozen. & \textbf{New} & \textbf{Confirmed} \\
69 & XiangShan & Debug & 4.9.1 & \texttt{dcsr.cause} reports 3 instead of 7 when a critical error coincides with a debug interrupt, and \texttt{tinfo.version} is 0 although \texttt{mcontrol6} is implemented. & Dup & Fixed \\
70 & XiangShan & Debug & 5.3 & When several triggers fire at once, the hart takes the breakpoint exception by trigger index instead of entering Debug Mode. & \textbf{New} & \textbf{Fixed} \\
71 & XiangShan & Debug & 5.5.4 & Address triggers compare only the low \texttt{VAddrBits} bits of the address, and \texttt{tdata2} is not legalized on write, so a non-canonical \texttt{tdata2} matches spuriously. & \textbf{New} & Reported \\
72 & XiangShan & Debug & 5.7.12 & The CBO store trigger drops \texttt{mcontrol6.match} and compares cache-block indices only, so GE and LT stop working; vector unit-stride accesses take the same equality path. & \textbf{New} & \textbf{Fixed} \\
73 & XiangShan & Debug & 5.7.12 & Writing \texttt{mcontrol6.chain} omits the required preceding \texttt{dmode} check. & \textbf{New} & \textbf{Fixed} \\
\end{xltabular}
}
  
\section{Audit Scope}
\label{app:scope}

The scope of Table~\ref{tab:specs} keeps the sections that specify the extensions a CPU supports. We take the supported extensions of CVA6 from its configuration packages~\cite{cva6repo} and those of XiangShan from its user guide~\cite{xiangshanuserguide}.

According to publicly available information in the XiangShan official repository, the XiangShan team plans to introduce a new implementation of the V extension, and its code has not yet been pushed to the repository. Therefore, this paper leaves the chapter out. We will audit it once the new implementation is merged.

\section{Cost by Step}
\label{app:cost}

Tables~\ref{tab:cost-detect} and~\ref{tab:cost-audit} break the cost
of Table~\ref{tab:cost} down by step: the two steps of implementation
localization and the two parts of candidate identification, and the
two reviews of the audit.

\begin{table}[h]
\centering
\caption{Cost of implementation localization and candidate
identification per campaign. \emph{Seed search} and \emph{File slice}
are the two steps of localization (Section~\ref{sec:localize});
\emph{Decision} and \emph{Retrieval} are the deciding model and the
retrieval sub-agents of identification (Section~\ref{sec:identify}),
the latter run only when the decision is \emph{not enough code}. Token
counts are in millions, input~\textbar~output; output tokens include
reasoning. \emph{Retr.}: share of rules for which retrieval ran.
\emph{Per rule}: mean time a rule takes from seed search to its final
decision.}
\label{tab:cost-detect}
\small
\setlength{\tabcolsep}{3pt}
\begin{tabular}{@{}l r P P P P P r r r r r@{}}
\toprule
& & \multicolumn{4}{c}{\textbf{Localization}} &
\multicolumn{4}{c}{\textbf{Identification}} & \multicolumn{2}{c}{} & & &
& \multicolumn{2}{c}{\textbf{Cost (\$)}} \\
\cmidrule(lr){3-6} \cmidrule(lr){7-10} \cmidrule(l){16-17}
\textbf{Campaign} & \textbf{Rules} &
\multicolumn{2}{c}{\hd{Seed search}{\inout}} &
\multicolumn{2}{c}{\hd{File slice}{\inout}} &
\multicolumn{2}{c}{\hd{Decision}{\inout}} &
\multicolumn{2}{c}{\hd{Retrieval}{\inout}} &
\multicolumn{2}{c}{\hd{\textbf{All steps}}{\inout}} &
\textbf{Calls} & \hd{\textbf{Retr.}}{\textbf{(\%)}} &
\hd{Per rule}{(min)} & Total & \hd{Per}{rule} \\
\midrule
\textit{CVA6} \\
\quad Unpriv & 1{,}212 & 27.5 & 93.0 & 51.4 & 118.1 & 8.0 & 39.3 & 46.0 & 13.0 & 132.8 & 263.3 & 11{,}808 & 16.8 & 20.5 & 186.5 & 0.154 \\
\quad Priv & 1{,}188 & 26.0 & 84.8 & 66.8 & 89.4 & 7.3 & 35.3 & 41.7 & 15.4 & 141.8 & 225.0 & 11{,}378 & 22.0 & 19.2 & 160.5 & 0.135 \\
\textit{XiangShan} \\
\quad Unpriv & 1{,}310 & 36.2 & 113.5 & 58.1 & 111.6 & 9.3 & 44.9 & 87.6 & 38.8 & 191.1 & 308.9 & 18{,}806 & 45.0 & 23.2 & 223.0 & 0.170 \\
\quad Priv & 1{,}431 & 39.5 & 99.3 & 55.9 & 126.3 & 8.7 & 47.6 & 81.5 & 31.6 & 185.7 & 304.8 & 18{,}315 & 29.8 & 19.6 & 220.3 & 0.154 \\
\quad AIA & 187 & 4.9 & 12.8 & 4.9 & 13.9 & 1.1 & 7.3 & 9.2 & 3.9 & 20.0 & 37.9 & 2{,}153 & 26.2 & 20.3 & 27.1 & 0.145 \\
\quad Debug & 357 & 10.8 & 29.2 & 11.6 & 30.4 & 1.8 & 9.4 & 12.7 & 4.4 & 36.9 & 73.4 & 3{,}988 & 23.5 & 19.1 & 52.6 & 0.147 \\
\midrule
Total & \nrules{} & 144.8 & 432.5 & 248.6 & 489.8 & 36.2 & 183.8 & 278.8 & 107.1 & 708.3 & 1{,}213.2 & 66{,}448 & 28.4 & 20.5 & 870.0 & 0.153 \\
\bottomrule
\end{tabular}
\end{table}

\begin{table}[h]
\centering
\caption{Cost of the two reviews per campaign. The
specification-side review runs on every candidate violation and the
code-side review on the candidates it keeps (\emph{Cand.} under each
review). Token counts are in millions, input~\textbar~output; output
tokens include reasoning. \emph{Rounds} and \emph{Time} are means per
candidate; a round is one model call in the review's search loop.
\emph{Per cand.} is the cost of both reviews per candidate entering the
audit.}
\label{tab:cost-audit}
\small
\setlength{\tabcolsep}{3pt}
\begin{tabular}{@{}l r r P r r r r r P r r r r r@{}}
\toprule
& \multicolumn{7}{c}{\textbf{Specification-side review}} &
\multicolumn{7}{c}{\textbf{Code-side review}} &
\multicolumn{2}{c}{\textbf{Both reviews}} \\
\cmidrule(lr){2-8} \cmidrule(lr){9-15} \cmidrule(l){16-17}
\textbf{Campaign} &
Cand. & Calls & \multicolumn{2}{c}{\hd{Tokens (M)}{\inout}} & Rounds & \hd{Time}{(s)} & \hd{Cost}{(\$)} &
Cand. & Calls & \multicolumn{2}{c}{\hd{Tokens (M)}{\inout}} & Rounds & \hd{Time}{(s)} & \hd{Cost}{(\$)} &
\hd{Cost}{(\$)} & \hd{Per cand.}{(\$)} \\
\midrule
\textit{CVA6} \\
\quad Unpriv & 215 & 1{,}565 & 21.0 & 5.0 & 7.3 & 165 & 3.96 & 92 & 1{,}406 & 26.0 & 5.6 & 15.3 & 463 & 4.44 & 8.41 & 0.039 \\
\quad Priv & 289 & 1{,}994 & 26.0 & 5.7 & 6.9 & 145 & 4.94 & 142 & 2{,}445 & 45.8 & 9.6 & 17.2 & 516 & 7.65 & 12.58 & 0.044 \\
\textit{XiangShan} \\
\quad Unpriv & 136 & 895 & 10.7 & 2.7 & 6.6 & 150 & 2.27 & 29 & 531 & 14.1 & 2.0 & 18.3 & 571 & 1.70 & 3.96 & 0.029 \\
\quad Priv & 300 & 2{,}127 & 27.6 & 5.6 & 7.1 & 159 & 4.89 & 79 & 1{,}379 & 30.2 & 4.9 & 17.5 & 539 & 4.09 & 8.98 & 0.030 \\
\quad AIA & 48 & 321 & 3.8 & 1.0 & 6.7 & 156 & 0.83 & 14 & 290 & 7.4 & 1.4 & 20.7 & 773 & 1.09 & 1.92 & 0.040 \\
\quad Debug & 95 & 742 & 10.2 & 1.9 & 7.8 & 154 & 1.69 & 28 & 525 & 11.6 & 1.8 & 18.8 & 534 & 1.48 & 3.17 & 0.033 \\
\midrule
Total & \ncand{} & 7{,}644 & 99.2 & 21.8 & 7.1 & 155 & 18.58 & \nspeckept{} & 6{,}576 & 135.0 & 25.2 & 17.1 & 523 & 20.45 & 39.02 & 0.036 \\
\bottomrule
\end{tabular}
\end{table}

\end{document}